\documentclass[twoside,twocolumn,10pt]{article}
\usepackage{extsizes}
\usepackage[super,sort&compress,comma]{natbib} 
\usepackage[version=3]{mhchem}
\usepackage[left=1.5cm, right=1.5cm, top=1.785cm, bottom=2.0cm]{geometry}
\usepackage{balance}
\usepackage{widetext}
\usepackage{times,mathptmx}
\usepackage{sectsty}
\usepackage{graphicx} 
\usepackage{lastpage}
\usepackage[format=plain,justification=raggedright,singlelinecheck=false,font={stretch=1.125,small,sf},labelfont=bf,labelsep=space]{caption}
\usepackage{float}
\usepackage{fancyhdr}
\usepackage{fnpos}
\usepackage[english]{babel}
\usepackage{array}
\usepackage{droidsans}
\usepackage{charter}
\usepackage[T1]{fontenc}
\usepackage[usenames,dvipsnames]{xcolor}
\usepackage{setspace}
\usepackage[compact]{titlesec}

\usepackage{epstopdf}

\definecolor{cream}{RGB}{222,217,201}

\begin{document}

\pagestyle{fancy}
\thispagestyle{plain}
\fancypagestyle{plain}{

\fancyhead[C]{\includegraphics[width=18.5cm]{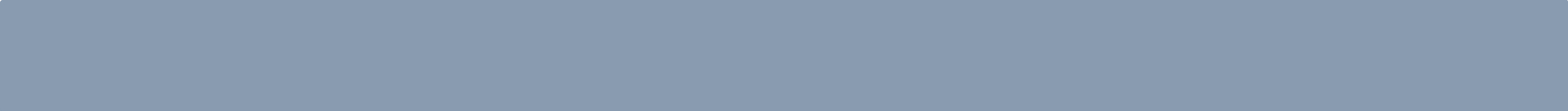}}
\fancyhead[L]{\hspace{0cm}\vspace{1.5cm}\includegraphics[height=30pt]{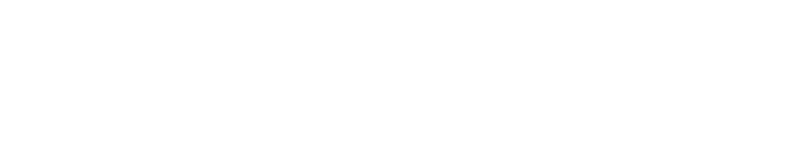}}
\fancyhead[R]{\hspace{0cm}\vspace{1.7cm}\includegraphics[height=55pt]{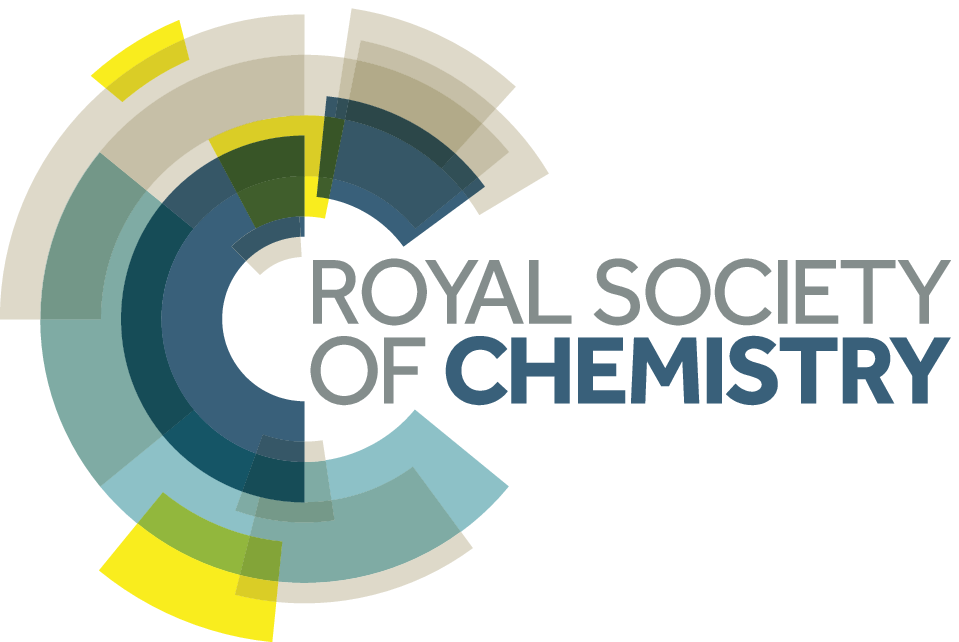}}
\renewcommand{\headrulewidth}{0pt}
}

\makeFNbottom
\makeatletter
\renewcommand\LARGE{\@setfontsize\LARGE{15pt}{17}}
\renewcommand\Large{\@setfontsize\Large{12pt}{14}}
\renewcommand\large{\@setfontsize\large{10pt}{12}}
\renewcommand\footnotesize{\@setfontsize\footnotesize{7pt}{10}}
\makeatother

\renewcommand{\thefootnote}{\fnsymbol{footnote}}
\renewcommand\footnoterule{\vspace*{1pt}%
\color{cream}\hrule width 3.5in height 0.4pt \color{black}\vspace*{5pt}} 
\setcounter{secnumdepth}{5}

\makeatletter 
\renewcommand\@biblabel[1]{#1}            
\renewcommand\@makefntext[1]%
{\noindent\makebox[0pt][r]{\@thefnmark\,}#1}
\makeatother 
\renewcommand{\figurename}{\small{Fig.}~}
\sectionfont{\sffamily\Large}
\subsectionfont{\normalsize}
\subsubsectionfont{\bf}
\setstretch{1.125} 
\setlength{\skip\footins}{0.8cm}
\setlength{\footnotesep}{0.25cm}
\setlength{\jot}{10pt}
\titlespacing*{\section}{0pt}{4pt}{4pt}
\titlespacing*{\subsection}{0pt}{15pt}{1pt}

\fancyfoot{}
\fancyfoot[LO,RE]{\vspace{-7.1pt}\includegraphics[height=9pt]{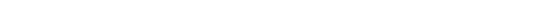}}
\fancyfoot[CO]{\vspace{-7.1pt}\hspace{13.2cm}\includegraphics{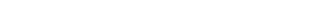}}
\fancyfoot[CE]{\vspace{-7.2pt}\hspace{-14.2cm}\includegraphics{head_foot/RF}}
\fancyfoot[RO]{\footnotesize{\sffamily{1--\pageref{LastPage} ~\textbar  \hspace{2pt}\thepage}}}
\fancyfoot[LE]{\footnotesize{\sffamily{\thepage~\textbar\hspace{3.45cm} 1--\pageref{LastPage}}}}
\fancyhead{}
\renewcommand{\headrulewidth}{0pt} 
\renewcommand{\footrulewidth}{0pt}
\setlength{\arrayrulewidth}{1pt}
\setlength{\columnsep}{6.5mm}
\setlength\bibsep{1pt}

\makeatletter 
\newlength{\figrulesep} 
\setlength{\figrulesep}{0.5\textfloatsep} 

\newcommand{\topfigrule}{\vspace*{-1pt}%
\noindent{\color{cream}\rule[-\figrulesep]{\columnwidth}{1.5pt}} }

\newcommand{\botfigrule}{\vspace*{-2pt}%
\noindent{\color{cream}\rule[\figrulesep]{\columnwidth}{1.5pt}} }

\newcommand{\dblfigrule}{\vspace*{-1pt}%
\noindent{\color{cream}\rule[-\figrulesep]{\textwidth}{1.5pt}} }

\makeatother

\twocolumn[
  \begin{@twocolumnfalse}
\vspace{3cm}
\sffamily
\begin{tabular}{m{4.5cm} p{13.5cm} }

\includegraphics{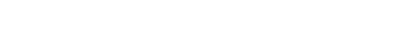} & \noindent\LARGE{\textbf{Ultracold Molecular Assembly}} \\
\vspace{0.3cm} & \vspace{0.3cm} \\

 & \noindent\large{Lee R. Liu,\textit{$^{abc}$} Jessie T. Zhang,\textit{$^{abc}$} Yichao Yu,\textit{$^{abc}$} Nicholas R. Hutzler,\textit{$^{abc}$} Yu Liu,\textit{$^{abc}$}   Till Rosenband,\textit{$^{b}$} and Kang-Kuen Ni$^{\ast}$\textit{$^{abc}$}} \\

\includegraphics{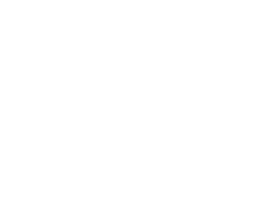} & \noindent\normalsize{Chemical reactions can be surprisingly efficient at ultracold temperatures ( $<1\,\mu$K) due to the wave nature of atoms and molecules. The study of reactions in the ultracold regime is a new research frontier enabled by cooling and trapping techniques developed in atomic and molecular physics. In addition, ultracold molecular gases that offer diverse molecular internal states and large electric dipolar interactions are sought after for studies of strongly interacting many-body quantum physics. Here we propose a new approach for producing ultracold molecules in the absolute internal and motional quantum ground state, where single molecules are assembled one by one from individual atoms. The scheme involves laser cooling, optical trapping, Raman sideband cooling, and coherent molecular state transfer.  As a crucial initial step, we demonstrate quantum control of constituent atoms, including 3D ground-state cooling of a single Cs atom, in a simple apparatus. As laser technology advances to shorter wavelengths, additional atoms will be amenable to laser-cooling, allowing more diverse, and eventually more complex, molecules to be assembled with full quantum control.  } \\

\end{tabular}

 \end{@twocolumnfalse} \vspace{0.6cm}

  ]

\renewcommand*\rmdefault{bch}\normalfont\upshape
\rmfamily
\section*{}
\vspace{-1cm}


\footnotetext{\textit{$^{a}$~Department of Chemistry and Chemical Biology, Harvard University, Cambridge, Massachusetts, 02138, USA.}}
\footnotetext{\textit{$^{b}$~Department of Physics, Harvard University, Cambridge, Massachusetts, 02138, USA.}}
\footnotetext{\textit{$^{c}$~Harvard-MIT Center for Ultracold Atoms, Cambridge, Massachusetts, 02138, USA.}}
\footnotetext{$^{\ast}$\textit{ E-mail: ni@chemistry.harvard.edu}}
	



\section{Introduction}
Ultracold molecules offer exciting new opportunities to explore the fundamental interface of chemistry and physics. On the one hand, molecules prepared in a pure internal and external quantum state allow studies of chemical reactions fully in the quantum regime~\cite{Krems2008}. Such studies will be of fundamental importance in chemistry. On the other hand, the rich molecular internal structure and interactions allow building systems that exhibit novel quantum many-body phases~\cite{Baranov2012}. Understanding and controlling fundamental chemical processes and physical interactions are central to  deciphering complex systems. 

Inspired by scientific and technical advances in the field of ultracold atomic and molecular physics, we propose a new approach to assemble ultracold molecules atom-by-atom. 
The novel source of single gas-phase molecules will  provide a new paradigm to study chemical reactions. By preparing exactly the required number of molecules for each experimental measurement, we could unambiguously determine the species involved in a reaction. Furthermore, by combining knowledge of the initial quantum state of the reactants and the detected quantum states of the reaction products, one can learn details of the chemical reactions dynamics and elucidate the processes of bond breaking and formation. Finally, a growing number of atoms in the periodic table can be laser-cooled, which means that the first step in the proposed approach - laser cooling of atoms to ultracold temperatures - is becoming increasingly applicable.  This new way of performing chemical reactions will enable the production and investigation of molecular species that were otherwise inaccessible.

Another motivation for creating a source of single molecules is its built-in individual molecule detection and manipulation capabilities which are crucial for applications in quantum information processing and quantum simulation. Quantum computers would allow new types of calculations for chemistry~\cite{AspuruGuzik2005}, opening the door to quantum-computing-based discovery of molecules for medicine, energy storage, and other practical uses. Single ultracold molecules can serve as quantum bits where information can be encoded in the long-lived hyperfine or rotational states while logic gate operations utilize their  long-range, tunable, and anisotropic interactions~\cite{Demille2002, Yelin2006}. The single molecules can also be interfaced with strip-line microwave cavities for quantum information applications~\cite{Andre2006}. Furthermore, the same dipole-dipole interactions that give rise to gate operations can be utilized to study novel quantum phases once the system is scaled up  to many molecules in a spatial configuration of one's choosing.

Many methods for trapping and cooling molecules to cold and ultracold temperatures have been demonstrated or are being pursued~\cite{carr2009}. We follow the general approach of assembling ultracold molecules from ultracold atoms. 
Complete internal quantum-state control of  molecules was achieved in bulk gases and optical lattices~\cite{Ni2008, Danzl2008, Lang2008, Chotia2012, Takekoshi2014, Molony2014, Park2015, Guo2016}. These advances have already opened the door to many research directions, including ultracold chemistry~\cite{Ospelkaus2010} and simulations of quantum spin models~\cite{Yan2013}. Furthermore, interesting strongly interaction phenomena can be explored with a lower entropy gas and with single molecule addressability~\cite{Moses2016, Gadway2016}. The method proposed here aims to realize both of these capabilities.

\section{The steps of single molecule production}

\begin{figure}[!t]
    \includegraphics[width=\columnwidth]{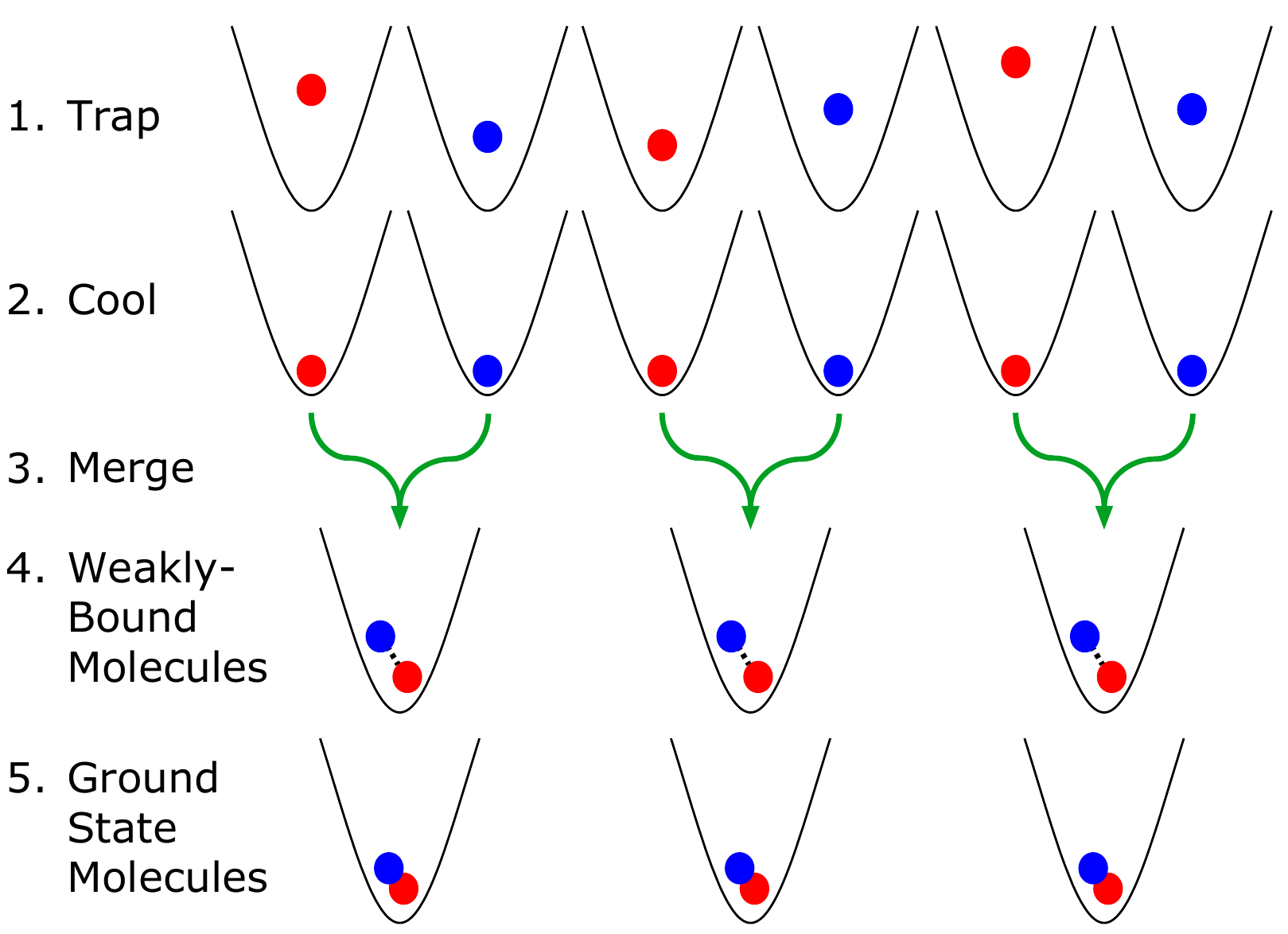}
  \caption{Step-by-step procedure for the ``ultracold molecular assembler.'' 1. Trap single atoms in an array of optical tweezer traps. 2. Cool atoms into motional ground state. 3. Merge a pair of  tweezers into one. 4. Convert atom pairs into weakly-bound molecules. 5. Perform coherent internal state transfer to bring weakly-bound molecules to the rovibronic ground state. }
\label{steps}
\end{figure}

Here we outline the steps  for single molecule production (``ultracold molecular assembler,'' see Fig.~\ref{steps}). This approach relies heavily on high fidelity internal and external quantum-state control of atoms and molecules and are generally applicable for a variety of species. Specifically, we choose diatomic molecules made from sodium (Na) and cesium (Cs) atoms due to their large molecular fixed-frame dipole moments (4.6 Debye)~\cite{Deiglmayr2008, Dagdigian1972} and extensive available spectroscopy~\cite{Grochola2011,Wakim2011,Zabawa2011}.  For ultracold chemical reaction studies, NaCs molecules prepared in different internal states offer distinct energetic pathways to participate in chemical reactions that could be switched on and off~\cite{Hutson2010}.

\begin{figure}[!t]
    \includegraphics[width=\columnwidth]{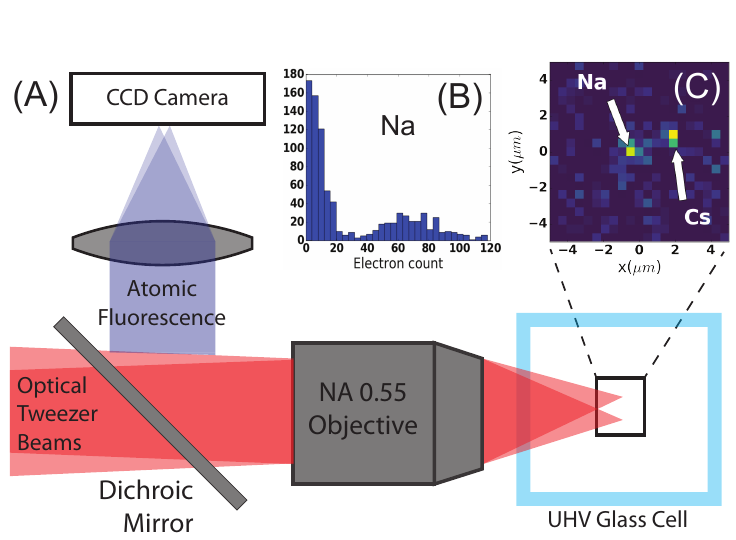}
  \caption{Single atom trapping and imaging setup and results. (A) Sketch of the apparatus. Optical tweezer beams are transmitted through a dichroic and focused by the objective into a glass cell. Fluorescence from trapped Na and Cs atoms is collected by the objective, reflected by the dichroic, and focused by a lens onto the CCD camera. (B) Histogram of photoelectron counts from the pixels in the CCD image of (C) corresponding to the Na atom. The doubly-peaked structure corresponds to a tweezer with zero or one atom. (C) Single-shot fluorescence image of single Na and Cs atoms trapped in adjacent tweezers.}
\label{app}
\end{figure}

The first step of the production (Fig.~\ref{steps}) is to prepare laser-cooled Na and Cs atoms in overlapped magneto-optical traps (MOTs) in a single vacuum chamber. A schematic of the apparatus is shown in Fig.~\ref{app}. A more detailed description can be found elsewhere~\cite{Hutzler2016}. The MOT serves as a cold atom reservoir for loading single atoms into tightly focused optical tweezer traps~\cite{Schlosser2001}. When a sufficiently  confined atom is illuminated with near-resonant light, a collisional blockade induces parity projection which limits the atom number to either zero or one. Loading a single atom therefore succeeds  approximately 50\% of the time due to its stochastic nature~\cite{Schlosser2002}.   To independently control Na and Cs,  we choose two different color tweezers, 970 nm for Cs ($\lambda_\text{D2}=852$ nm) and 700 nm for Na ($\lambda_\text{D2}=589$ nm). The severe light shifts of Na at the trapping wavelength would normally prevent atom cooling, and consequently, efficient atom loading. We eliminate the light shift through fast temporal alternation of trapping (tweezer) and cooling~\cite{Hutzler2016}. This technique should enable loading of other species of atoms and molecules that might otherwise experience large light shifts. Subsequently, fluorescence imaging determines if a single atom has been successfully loaded before proceeding. Simultaneous trapping of single Na and Cs atoms side-by-side has been achieved, as shown in  Fig. \ref{app}.

To maximize wavefunction overlap between two atoms so that they can be efficiently converted into a molecule, we desire atoms trapped in the same tweezer and with the smallest possible wavefunction spread. To achieve this, the second experimental step (Fig.~\ref{steps}) is to cool the atoms into their motional ground state  by applying 3D Raman sideband cooling (RSC), a technique first demonstrated with single ions~\cite{Monroe1995} and recently with single neutral Rb atoms~\cite{Kaufman2012, Thompson2013}. Once Na and Cs atoms are in their 3D motional ground state,  the two tweezers will be overlapped and merged. The Na tweezer trap will be adiabatically ramped down to ensure that both Cs and Na are trapped at the same position. 

The smallest wavefunction spread achievable in the confining tweezer quadratic potential corresponds to the zero-point length, $z_0 = \sqrt{\hbar/(2 m\omega)}$, where $m$ is the atomic mass, $\hbar$ is the reduced Planck's constant, and $\omega$ is the angular frequency of the trapped atom. For a representative $2\pi\times70$ kHz trap frequency, the zero-point lengths for Na and Cs are $z_0=1059\,a_0$ and $z_0=440\,a_0$ respectively, where $a_0 = 0.053$ nm is the Bohr radius. Therefore, two atoms cooled to their motional ground state will still have an extent in one dimension more than a hundred times larger than a typical molecular bond length, making efficient conversion challenging.

Following RSC, we will merge the two tweezers (Fig.~\ref{steps}) in order to initialize the Na and Cs atoms in a single internal and motional quantum state. In a quadratic potential, as in the case of the bottom of an optical tweezer, two non-interacting particles can be recast into relative and center-of-mass coordinates~\cite{Saenz2007}. Two atoms prepared in their respective motional quantum ground states can then be viewed as being in a single relative motional state. A pair of atoms can then be converted to a molecule in its absolute rovibrational ground state with almost unit efficiency via quantum state transfer, provided that the entire process is coherent. 

To overcome technical challenges associated with bridging two quantum states with a large energy difference but small wavefunction overlap, we plan to use two steps (Fig.~\ref{steps}). A sodium and a cesium atom in the same tweezer will  be first associated into the most weakly bound molecular state (vibrational level $\nu''=24$ in $a^3\Sigma^+$) by a detuned, coherent, two-photon Raman pulse. Subsequently, the weakly-bound molecule will be transferred into the rotational, vibrational, and electronic (rovibronic) ground state by STImulated Raman Adiabatic Passage (STIRAP)~\cite{STIRAP}. To check that the process is successful, residual atoms will be blown out with resonant light, and the surviving molecule will be coherently converted back into atoms for fluorescence imaging.

In the following sections, we will detail the steps of motional and internal state control of molecules with experimental demonstrations and theoretical calculations.

\section{Controlling the quantized motion of atoms}

Using standard MOT and polarization gradient cooling (PGC), we can cool a single Cs or Na atom to the point where the discreteness of the trap's energy levels becomes important. 
To further manipulate the atom into the lowest motional state, it is necessary to operate in the resolved sideband regime where the linewidth of the cooling transition is less than the trap frequency (10-100's of kHz). Raman sideband cooling (RSC) has been demonstrated for a variety of systems in this regime~\cite{Monroe1995, Teufel2011, Kaufman2012}.  Here, we demonstrate RSC of a single Cs atom to its 3D motional ground state. 

\begin{figure}[!t]
      \includegraphics[width=\columnwidth]{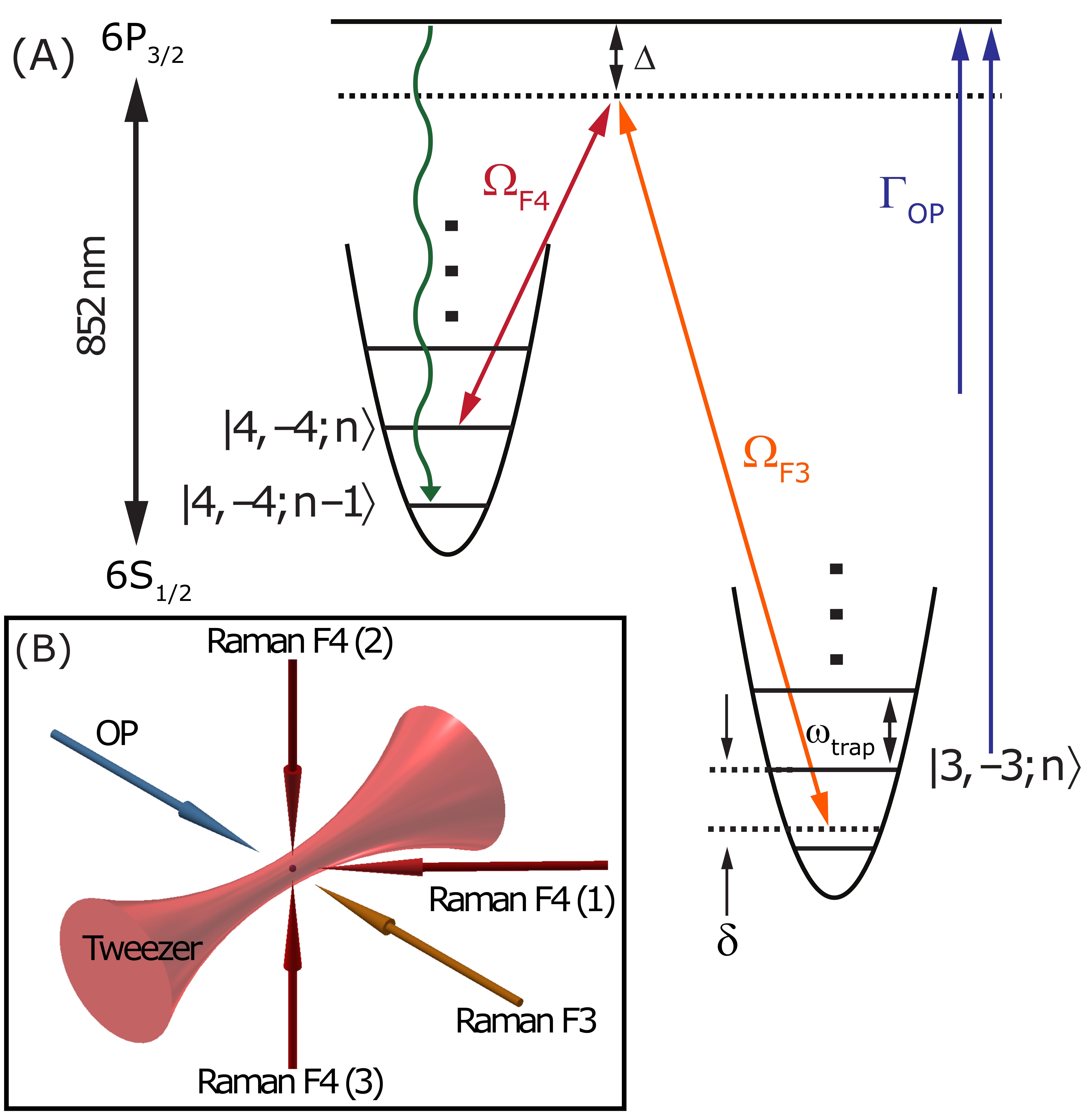}
  \caption{Schematic of Raman sideband cooling (RSC) in Cs. (A) Energy level diagram depicting the states involved. $\Omega_\text{F3,F4}$, Rabi frequency of Raman beams; $\Gamma_\text{OP}$, optical pumping rate; $\Delta$, one-photon detuning; $\delta$, two-photon detuning, difference between the F3 and F4 laser frequencies minus the hyperfine splitting; $n$, initial motional quantum number.  Internal and external quantum state denoted in the form $|F,m_F;\nu\rangle$. (B) Physical setup of laser beams. We cool all 3 axes (2 radial, 1 axial) independently by keeping Raman F3 direction constant while toggling the direction of the Raman F4 beams. OP refers to the optical pumping beam.}
\label{rsc_scheme}
\end{figure}

The RSC sequence consists of two steps: a coherent two-photon Raman transition that connects two internal states while removing a motional quantum, and an optical pumping (OP) step that re-initializes the internal state of the atom. The two steps are repeated until the atom population reaches the ground state. Although we will discuss cooling of an atom initialized in a single motional state $\nu=n$ where $\nu$ is the motional quantum number, the following mechanism also applies to distributions of motional states, as in the case of a thermal state prepared by PGC. 

In our scheme (Fig.~\ref{rsc_scheme}),  the Raman transition is  between Cs ground-state hyperfine levels   $|F,m_\text{F}\rangle=|4,-4\rangle$ and $|3,-3\rangle$, which are about 9.2 GHz apart. To drive this transition, we phase-lock two diode lasers and set the one-photon red detuning $\Delta=2\pi\times44$ GHz from the Cs $D_2$ line at 852 nm.
To further address the motional degrees of freedom, the laser beams are arranged as shown in Fig.~\ref{rsc_scheme}(B) to achieve substantial two-photon momentum transfer, $\Delta \vec{k} = \vec{k}_\text{F4(i)}-\vec{k}_\text{F3}$, while the energy difference associated with the hyperfine level and motional state change is supplied by the two-photon detuning, $\delta$. If the bottom of the trap is sufficiently harmonic, as is the case here, this resonance condition is maintained for all relevant motional states,  $\nu$. 

The atom is initially prepared in $|4,-4\rangle$ by OP. We use $\sigma^{-}$-polarized beams resonant with $|4,-3\rangle \rightarrow |4',-4\rangle$ and $|3,-3\rangle \rightarrow |4',-4\rangle$ transitions, where the primed levels denote sub levels of $6P_{3/2}$. During the first step of RSC, a Raman $\pi$-pulse is applied between $|F,m_\text{F};\nu\rangle=|4,-4;n\rangle$ and $|3,-3; n-1\rangle$.  Subsequently, OP is pulsed on to cycle the atom back to $|4,-4\rangle$. 
The OP step has a high probability of preserving the motional state~\cite{wineland1998}. Thus, in one RSC cycle, $\nu$ has decreased by nearly one on average. This process repeats until the atom reaches the dark state $|4,-4;0\rangle$, thereby deterministically preparing the internal and the motional quantum state of the atom. 
To address each axis separately, we toggle the three Raman F4($i$) beams by repeating the sequence ${i=3,1,2,1}$. The linewidth of the Raman transition is $\pi$-time Fourier limited due to the effective Raman coupling  $\Omega_\text{R}=\Omega_\text{F3}\Omega_\text{F4}/2\Delta =2\pi\times33$kHz ($2\pi\times7\,$kHz) for radial (axial) trap axes.  OP with a rate of $\Gamma_\text{OP}=2\pi\times83$ kHz is applied for $85\,\mu s$ in between cooling cycles. For the entirety of RSC, the quantization axis is defined by a 8.6 G magnetic field along the OP propagation direction. 

All Raman pulses in this experiment for cooling and spectroscopy use a Blackman window temporal intensity profile to reduce off-resonant excitation of the carrier. The starting temperature of $9.2\,\mu$K, corresponding to an initial average motional quantum number $\bar{n}_\text{a}=9$ in the loosely confined axial direction, leads to significant occupation (i.e., at least $1\%$ of the $n_\text{a}=0$ population) of levels up to $n_\text{a}=40$. Some of these higher-lying levels could not be cooled efficiently by only applying RSC pulses for the duration of a ground-state sideband $\pi$-pulse due to the $\sqrt{n}$ scaling of coupling of adjacent motional states~\cite{wineland1998}. Therefore, we sweep the axial pulse duration to address all levels in descending order from $n_\text{a}=n_\text{init}\rightarrow1$. To overcome decoherence, which reduces the transfer fidelity of each pulse, we repeat the sweep with decreasing $n_\text{init}=\left\{41,31,16,11,6\right\}$.
 
\begin{figure}[!t]
      \includegraphics[width=\columnwidth]{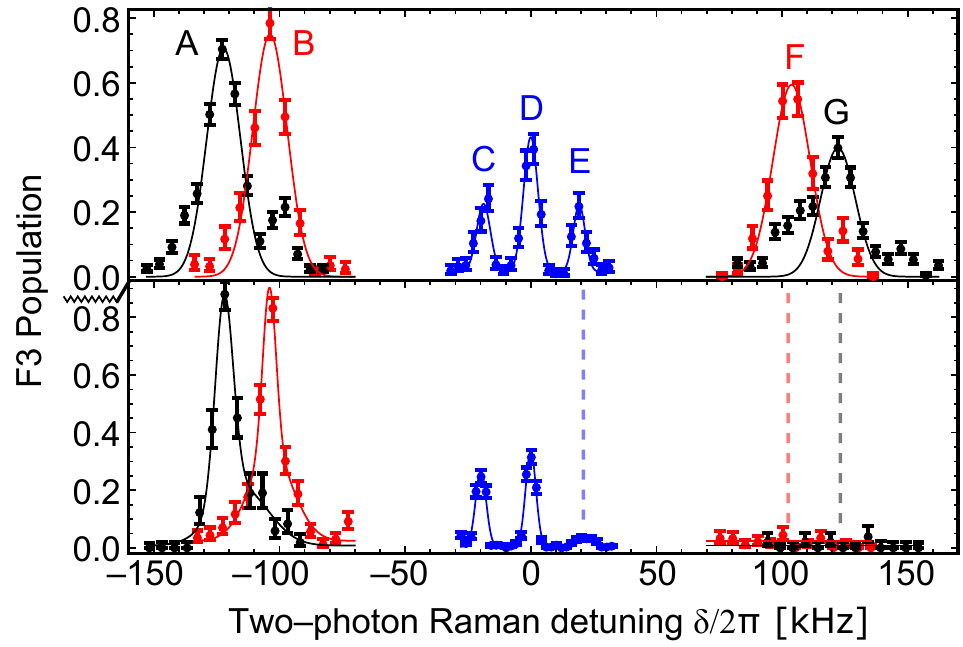}
  \caption{3D Raman sideband spectra for Cs. [Top] Before RSC. A, B, and C refer to heating sidebands for the two radial and  the axial directions. D refers to the axial carrier. E, F, and G refer to cooling sidebands for the axial, and the two radial  directions. The temperature is $9.2\mu$K, corresponding to a 3D motional ground state population of $P_{(0,0,0)} = 2.5\%$. [Bottom] After RSC is applied. Sideband thermometry indicates $P_{(0,0,0)}=84_{-5}^{+3}\%$. Solid lines are fits to independent Gaussians, and only serve to extract the center of the peaks.}
\label{RSCdata}
\end{figure}

To verify cooling performance, we compare Raman sideband spectroscopy (Fig.~\ref{RSCdata}) of the single atom with and without cooling. The atom population in the $F=3$ manifold was detected by blowing out atoms in the $F=4$ manifold with light resonant with the cycling $|4,-4\rangle \rightarrow |5',-5\rangle$ transition. The trap cross section is elliptical giving rise to different radial trap frequencies. The sidebands feature a main peak and a smaller side peak, the latter of which we attribute to interference between the sideband and the overdriven carrier. Nevertheless, equating the ratio of the heights of the main peaks of the two sidebands to $\frac{\bar{n}+1}{\bar{n}}$ should give the correct $\bar{n}$. This was verified by simulating sideband spectroscopy on a thermal atom using a master equation evolution of the density matrix. By assuming a thermal distribution, we can extract a temperature and a ground state probability for each axis. For the hot atom, we measure a temperature of $9.2\mu K$, corresponding to a 3D motional ground state population of $P_{(0,0,0)} = 2.5\%$. After RSC, $(\bar{n}_\text{a}, \bar{n}_\text{r1},\bar{n}_\text{r2}) = (0.14_{-0.03}^{+0.04},0.03_{-0.02}^{+0.03},0.01_{-0.01}^{+0.04})$, corresponding to a ground-state population $(P_{n_\text{a}=0},P_{n_\text{r1}=0},P_{n_\text{r2}=0}) = (87_{-3}^{+3},98_{-3}^{+2},99_{-3}^{+1})$. Hence, a 3D ground-state population of $P_{(0,0,0)}=84_{-5}^{+3}\%$ is reached  after 100 cycles of RSC ($\approx100$ ms total cooling time). 

The ground-state population achieved following RSC is primarily limited due to the fact that the choice of $\Omega_\text{R}$ is a trade-off between overcoming decoherence mechanisms while maintaining adequate sideband resolution. 
Efficient cooling requires $\eta_\text{R} \Omega_\text{R}>\frac{2\pi}{T_{2}}$, where in our experiment $\eta_\text{R} =0.17 (0.14)$  is the Lamb-Dicke parameter~\cite{wineland1998} for the axial(radial) direction and $T_{2} = 300\,\mu s$ is the coherence time of the Raman transition, limited mainly by external magnetic field fluctuations and the phase coherence of the Raman transition lasers. Future improvement may be achieved by increasing trap frequency, and therefore the upper bound on $\Omega_\text{R}$.

RSC will also be applied to prepare a single Na atom in its motional ground state. A sodium atom has a higher photon recoil frequency due to its lighter mass, which means that the heating rate due to OP is increased and that the atom is initially hotter following preparation by PGC. Initial RSC is therefore confounded by motional de-phasing of the hot atom and needs to overcome the increased heating rate due to OP.  However, the light mass also means the atom has a higher oscillation frequency in the trap relative to Cs, giving sidebands that are more highly resolved, and the larger $\eta_\text{R}$ allows coupling of significant higher order sidebands. Successful RSC of Na will therefore entail balancing the higher heating rates due to optical pumping with higher cooling rates afforded by large sideband separation and higher order sidebands.

\section{Weakly-bound molecules}
Once both Cs and Na atoms are prepared in their motional ground state, the atoms will be merged into the same tweezer. The final task in the ultracold molecular assembler (Fig.~\ref{steps}) is to convert them into a molecule that occupies the rovibronic ground state. We plan to achieve this goal via two separate coherent two-photon processes: first, converting atoms into a weakly bound molecule; second,  transferring the molecule from a weakly bound state to the rovibronic ground state. 
Many methods have been demonstrated to convert cold atoms to weakly-bound molecules~\cite{Wynar2000,Donley2002,Rom2004,Sage2005,Stellmer2012}, but a coherent method is necessary for high fidelity conversion. We numerically explore molecule formation utilizing a two-photon Raman pulse~\cite{Wynar2000, Jaksch2002, Rom2004} in a parameter regime distinct from other methods. 

\begin{figure}[!t]
      \includegraphics[width=\columnwidth]{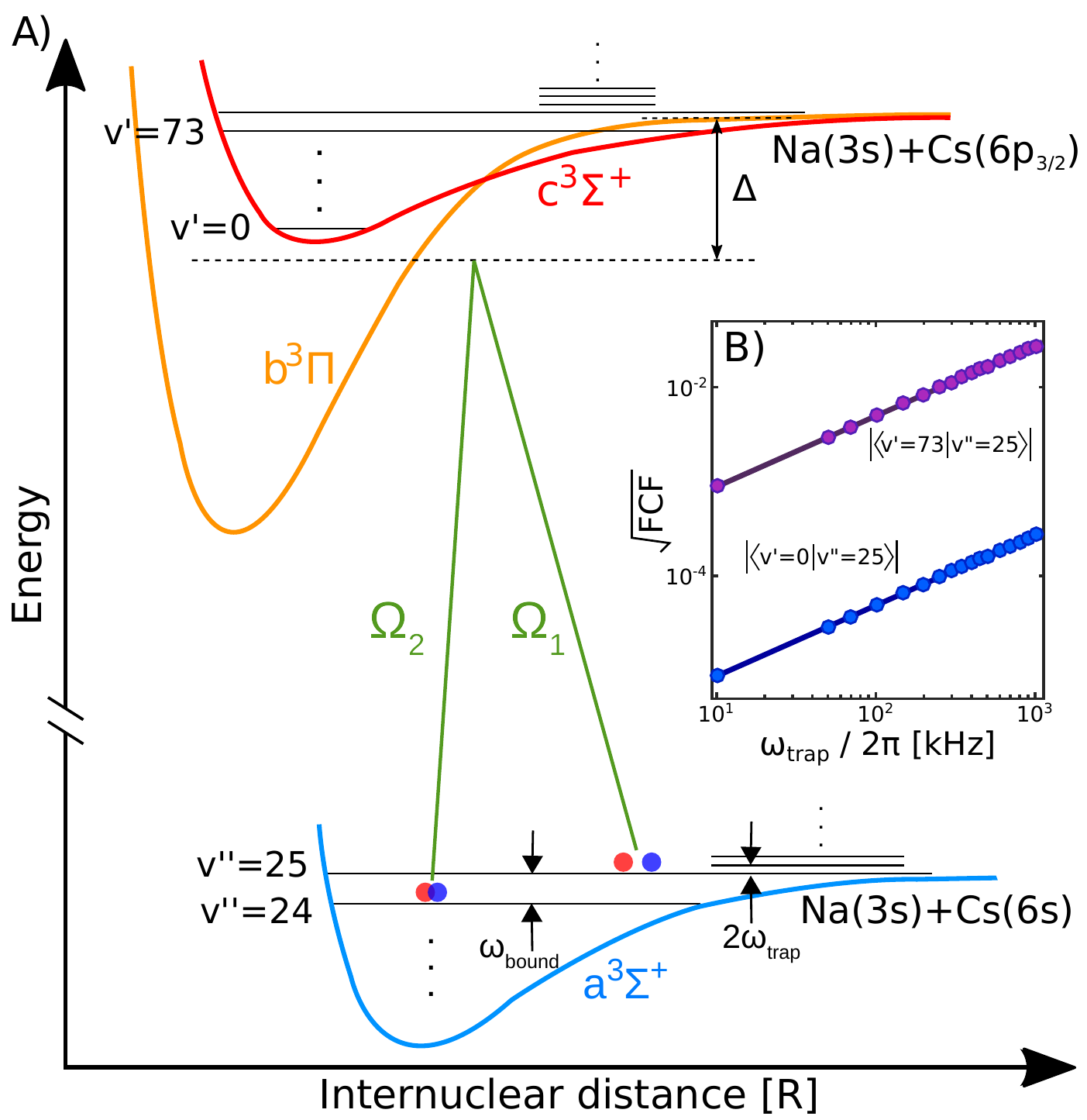}
  \caption{Weakly-bound molecule formation scheme. (A) Energy level diagram depicting states involved (not to scale). Raman transition lasers (green) couple the initial state $\nu''=25$ to the weakly bound molecular state $\nu''=24$ in the $a^3\Sigma^+$ potential. To minimize spontaneous emission, we tune the lasers $\Delta=\sim2\pi\times(-63)$ THz below the Na(3s)+Cs(6p$_{3/2}$) atomic asymptote.  (B) Magnitude of the wavefunction overlap  for $\nu''=25$ in $a^3\Sigma^+$ to $\nu'=73$ (purple) and to $\nu'=0$ (blue) in  $c^3\Sigma^+$ vs. trap frequency. Solid lines show fits corresponding to a scaling of  $\omega_\text{trap}^{3/4}$ in both cases.}
\label{weaklyboundscheme}
\end{figure}

The two-photon Raman transfer utilizes electronic transitions through an intermediate excited state and requires sufficiently large Franck-Condon factors (FCF) between all relevant levels in order to realize efficient conversion (Fig.~\ref{weaklyboundscheme}). We calculate all bound states of the ground $a^3\Sigma^+$ NaCs potential~\cite{Docenko2006} and the excited $c^3\Sigma^+$ and $b^3\Pi$ potentials~\cite{Grochola2011, Zaharova2009} using the mapped Fourier method~\cite{Fattal1996}. We include the tweezer confinement out to a distance of $2.5 \mu$m in our calculations, but make the simplifying assumption that the atoms have the same spherical harmonic trapping frequency.\footnote{This assumption is roughly true because the ratio of Na and Cs polarizabilities  at a trapping wavelength around $\lambda=1\,\mu$m to their masses  are similar~\cite{Safronova2006}. This condition is not an experimental requirement.} This assumption allows the Hamiltonian of two non-interacting atoms to be separated into their center-of-mass (CM) and relative (internuclear separation) coordinates~\cite{Saenz2007}, $R$. Because molecular conversion depends only on the relative motion of the two atoms, we can neglect the CM term and add the tweezer confinement, as $\frac{1}{2}\mu\omega_\text{trap}^2 R^2$, where $\mu$ is the reduced mass and $\omega_\text{trap}$ is the trap frequency, directly onto the molecular potentials. 

Tight  harmonic confinement gives the wavefunction appreciable amplitude inside the internuclear potential well for the unbound atomic vibrational levels (the lowest one starting at $\nu''=25$), and is needed to enhance the FCF between levels of unbound atoms to excited molecular states.  As an example, the magnitude of the wavefunction overlap ($\sqrt{\text{FCF}}$) of the transition between $\nu''=25$ to $\nu'=73$ and  to $\nu'=0$ of the excited molecular state as a function of trap frequency are shown in Fig.~\ref{weaklyboundscheme}(B). Our numerical enhancement scales with 
$\sqrt{\text{FCF}}\propto \omega_\text{trap}^{3/4}$, consistent with an analytical expression found previously~\cite{Mies2000}. 

\begin{figure}[!t]
      \includegraphics[width=\columnwidth]{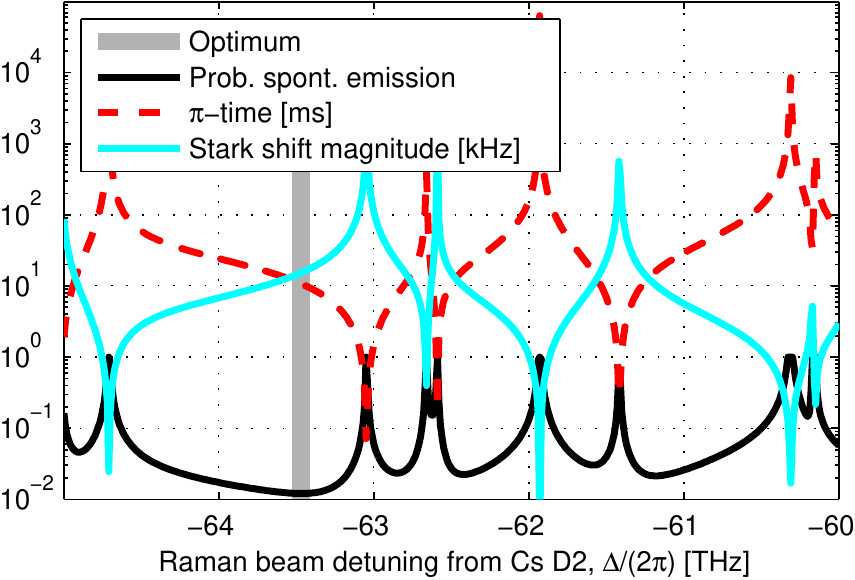}
  \caption{Probability of spontaneous emission (PSE), $\pi$-time (transfer duration) and differential Stark shift between the states $\nu''=25$ and $\nu''=24$ for the Raman beams, as a function of detuning from the Cs $D_2$ transition ($\Delta/(2\pi)$).  The beams are expected to have equal intensity (10 mW power, 10 $\mu$m beam waist).  The calculation includes vibrational levels $\nu''=24$ and $25$ of the $a^3\Sigma^+$ ground state and a complete basis of vibrational eigenstates derived from the $b^3\Pi$ and $c^3\Sigma^+$ excited state molecular potentials which are embedded in a harmonic well.  The dipole matrix elements are assumed to be $3ea_0$ times the relevant wavefunction overlap.  Minimum PSE (1.2~\%) occurs at $\Delta=2\pi\times-63.468$ THz where the $\pi$-time is $10.7$ ms.  The beams cause a total differential Stark shift of $15$ kHz, necessitating a laser intensity stability better than $0.5$ \%.}
\label{transfervsdet}
\end{figure}

Experimentally, we will couple the initial unbound atom pair ($\nu''=25$) to the least bound molecular state ($\nu''=24$) in the $a^3\Sigma^+$ potential via a two-photon Raman transition.  To drive this transition, the two Raman laser beams have a difference frequency of $\omega_\text{bound}\approx 2\pi\times 282 \text{MHz}$, the binding energy of the weakly bound $\nu''=24$ state. The choice of Raman detuning is critical to avoid excessive spontaneous emission and Stark shifts while maintaining a sufficiently high two-photon Rabi rate~\cite{Wineland2003}.  We show the calculated values for these parameters in Fig.~\ref{transfervsdet}, together with the expected operating point, where the Raman lasers are tuned several 100's of GHz below the lowest vibrational state of the excited-state $c^3\Sigma^+$ potential.  At this detuning, the Raman beams have a wavelength of 1037 nm, in a regime conveniently accessible by Ytterbium-doped fiber lasers and amplifiers.

Although the FCFs that connect weakly-bound molecular states to the harmonic oscillator ground state are enhanced by the harmonic confinement, the calculated transition still requires a relatively long pulse duration of $11$ ms throughout which the Raman beams must stay coherent.  However, the coherence requirement can easily be met by generating the two beams from a single laser with an acousto-optic modulator (AOM).  Because the expected pulse duration is significantly longer than the harmonic oscillator periods of the ODT, the Raman beams will not couple to other vibrational levels (e.g. $\nu''=26$) of the $a^3\Sigma^+$ potential.  We expect the dominant terms that limit transfer fidelity to be spontaneous emission (about 1 \%), incomplete ground-state cooling, and fluctuating Stark shifts (see Fig.~\ref{transfervsdet}).

\section{Ground-state molecules}
The final step of single molecule creation (Fig.~\ref{steps}) is to convert the molecule from a weakly-bound triplet ground state to the rovibronic singlet ground state (or potentially any other desired internal quantum state). This step  relies on identifying a suitable intermediate excited state that has favorable FCF's with the initial and target states, together with a large spin-orbit coupling, to transfer between triplet and singlet molecular states. Such a transfer scheme has been demonstrated efficiently ($\approx95\%$) for a number of ultracold bi-alkali dimers~\cite{Takekoshi2014,Molony2014,Park2015,Guo2016}. In particular, a wise choice of intermediate molecular excited state would require only modest laser powers for the two-photon STIRAP transfer beams. The main challenge of this step lies in the technical aspect of maintaining coherence of two vastly different wavelength lasers. Based on prior spectroscopy~\cite{Zabawa2011}, we expect to use laser wavelengths around 905 nm and 635 nm to drive a coherent STIRAP transition from the initial to the final state.  To maintain their coherence, both lasers can be simultaneously locked to a single stable high-finesse reference cavity~\cite{Aikawa2011}.  Thus, single molecule production after stochastic loading of individual atoms into optical tweezers is expected with a fidelity of $>65\%$.
To verify that a single molecule has been produced, we plan to  reverse the coherent production steps to convert the molecule back to atoms.  The dissociated atoms will then be separated into individual tweezers and imaged using fluorescence with 98\% fidelity.

\section{Conclusions}
We present a scheme to produce a novel source of single ultracold polar molecules.  Our scheme relies on high fidelity quantum state manipulation of atoms and molecules. We have  demonstrated the crucial first steps toward single molecule production, which includes trapping the constituent atoms (Cs and Na) side-by-side in optical tweezers and motional ground-state  cooling of Cs to its 3D ground state (84\%). We numerically explore a two-photon Raman scheme to convert atom pairs to a weakly-bound molecule by utilizing both an enhanced FCF due to the trapping confinement and  far-detuned Raman beams to minimize spontaneous emission. We further highlight the spectroscopy and technical requirements to transfer the weakly-bound molecule to its rovibronic ground-state. 

Although our initial effort concentrates on producing one molecule, the number of single molecules can be scaled up by employing an array of single atom tweezer traps as a starting point. Molecules will be assembled from  pairs of atoms in parallel by the same production described above. The optical tweezer array can be generated by a variety of technical tools such as digital mirror devices or by sending a single tweezer through an acousto-optical deflector driven by multiple radio-frequency tones. In such a 1D or 2D tweezer array, unity atom filling  can be achieved by real-time re-arrangement of tweezer locations~\cite{Barredo2016, Endres2016}, which  provides a promising starting point for a large, flexible, 2D array of  molecules. This ultracold molecular assembler features fast cycle time, flexible geometry, and individual control and imaging capabilities, and provides an attractive platform for studies of ultracold chemistry, quantum information, and many body physics. 

Note added: after completion of our work, we became aware of a related work demonstrating RSC of a single Cs atom~\cite{robens2016}.

\section*{Acknowledgments} 

We thank Cindy Regal for inspiring discussions. K.-K. Ni thanks Deborah Jin for  encouragement to pursue the research  presented here.  This work is supported by the NSF through the Harvard-MIT CUA, as well as the AFOSR Young Investigator Program, the Arnold and Mabel Beckman Foundation, and the Alfred P. Sloan Foundation.

\bibliography{refs} 
\bibliographystyle{rsc} 

\end{document}